\documentclass[twocolumn,apj]{emulateapj}

\shorttitle{GRB 060505 ISM Properties}
\shortauthors{Levesque \& Kewley}

\begin{document}

\title{The Host Galaxy of GRB 060505: Host ISM Properties}

\author{Emily M. Levesque and Lisa. J. Kewley}
\affil{Institute for Astronomy, University of Hawaii, 2680 Woodlawn Dr., Honolulu, HI 96822}
\email{emsque@ifa.hawaii.edu, kewley@ifa.hawaii.edu}

\begin{abstract}
We investigate the ISM environment of GRB 060505.    Using optical emission-line diagnostic ratios, we compare the ISM properties of the GRB 060505 host region with the hosts of unambiguous long- and short- duration GRBs.  We show that the metallicity, ionization state, and star formation rate of the GRB 060505 environment are more consistent with short-duration GRBs than with long-duration GRBs.  We compare the metallicity and star formation rates of the GRB 060505 region with four other star-forming regions within the GRB 060505 host galaxy.  We find no significant change in metallicity or star formation rate between the GRB 060505 region and the other four host regions.  Our results are consistent with a compact-object-merger progenitor for  GRB 060505.  
\end{abstract}

\keywords{gamma-ray bursts}

\section{Introduction}
\label{Sec-intro}
The nature of the GRB 060505 progenitor is currently a topic of hot debate.
GRBs are the signatures of extraordinarily high-energy events. Burst length
distinguishes between ``short'' ($<$ 2 s) bursts arising from compact-object mergers (Gehrels et al. 2005) and ``long'' ($>$ 2 s) bursts with massive core-collapse progenitors (Woosley 1993) that
are commonly accompanied by luminous and broad-lined Type Ic supernovae (Watson et al. 2007).
GRB 060505 has a burst length of $\sim$4 s, but notably lacks evidence of an accompanying supernova.   Investigations into the host properties of GRB 060505 strongly
disagree on the nature of the progenitor.   It is unclear whether GRB 060505 originates from 
a compact-object merger, a massive core-collapse supernova, or a new class of long-duration GRBs with no associated supernovae.   The nature of GRB 060505 may have important implications for our classification and understanding of GRB progenitors.

GRB 060505 was observed on UTC 2006 May 5 by the Swift Burst Alert Telescope (BAT)
(Hullinger et al. 2006; Palmer et al. 2006), associated with the $z$ = 0.0889 galaxy
2dFGRS S173Z112 (Colless et al. 2003; Ofek et al. 2006; Th\"{o}ne et al. 2006a; Fynbo et al. 2006).  It was initially categorized as a long-duration GRB based on its $\sim$4 s burst length (Kouveliotou et al. 1993).    Th\"{o}ne \& Fynbo (2007) find a lower metallicity and higher rate of star formation at the GRB 060505 burst site when compared with other regions of the host galaxy. Recent investigations suggest that long-duration GRBs are associated with low-metallicity star-forming environments (Stanek et al. 2006, Sollerman et al. 2005, Fruchter et al. 2006, Kewley et al. 2007, Brown et al. 2007), supporting a core-collapse progenitor scenario for GRB 060505.

On the other hand, GRB 060505 may be the product of a compact-object merger with a longer-than-average burst duration.  Short- and long-duration GRBs are separated by a burst-duration cut-off of 2 s, but there may be some overlap between these two classes of progenitor
events; short-burst progenitors have a 12\% chance of yielding a burst longer
than 4 s (Horv\'{a}th 2002).   

Additional support for a compact-object merger progenitor for GRB 060505 includes the progenitor's evolutionary timescale, the spiral nature of the host galaxy, and the brightness of the burst region.  Ofek et al. (2007) calculate an upper limit of 10 Myr for the progenitor birth-to-explosion timescale of the GRB 060505 event. While this age limit does not rule out the possibility of a core-collapse progenitor, such a timescale is also consistent with the merging of two neutron stars or a neutron star-black hole merger, both of which are compact object merger scenarios associated with short bursts (Belczynski et al. 2006).  The host galaxy of GRB 060505 is categorized as an Sbc spiral, which is unusual for a long-duration GRB host galaxy (Th\"{o}ne \& Fynbo 2007).   Fruchter et al. (2006) found that long-duration GRBs favor the brightest regions of their host galaxies that are associated with concentrated populations of young massive stars (van den Heuvel \& Yoon 2007).   The GRB 060505 progenitor region is relatively faint compared to its host galaxy, supporting a compact-object merger progenitor (Ofek et al. 2007).

Alternatively, GRB 060505 may belong to a new class of long-duration GRBs with
no associated supernovae.   The distribution of known GRB burst durations suggest the 
existence of a third category of GRBs (Mukherjee et al. 1998, Horv\'{a}th 2002).   
GRB 060505 is often compared with GRB 060614, a $\sim$102 s burst (Barthelmy et al. 2006) classified as a long GRB with no apparent supernova counterpart - both have been proposed as representative examples of a new class of GRBs (Fynbo et al. 2006, Jakobsson \& Fynbo 2007, King et al. 2007).

Schaefer \& Xiao (2006) suggest that GRB 060505 is a background event that has been associated with 2dFGRS S173Z112 by coincidence.  However, Watson et al. (2007) estimate that
the superposition of the burst directly over a star-forming region of low metallicity would be unreasonably serendipitous.

There are several reasons to believe that the progenitors of long-duration bursts
would favor low-metallicity environments. Mass loss in late-type massive
stars, driven by radiation pressure on spectral lines, is heavily dependent on
metallicity (Vink \& de Koter 2005), with mass loss and metallicity
correlated by the rough relation $\dot M_w \propto Z^{0.78}$ (Mokiem et al.
2006). Surface velocities are also expected to be higher for such stars at
low metallicity, a consequence of the lower mass loss rate and an important
property of collapsars (Kudritzki \& Puls 2000, Meynet \& Maeder 2005).
The host environments of long-duration GRBs should also have high
ionization parameters, since the typical age of the young stellar populations in
long-duration bursts is
consistent with late-type massive stars that dominate the radiation
field, such as Wolf-Rayet stars (Kewley et al. 2007).
There is no evidence that a compact-object-merger progenitor would favor such an environment: compact-object mergers are found in many galaxy types (such as
ellipticals and spirals, including early-type spirals,) that typically have older stellar
populations (and, therefore, less ionizing radiation) and considerably smaller
star formation rates (Nakar 2007). These hosts typically have
higher metallicities than the blue compact dwarf galaxy hosts of
core-collapse-progenitor long GRBs (Bloom \& Prochaska 2006). These characteristics
suggest that the ISM properties of these two host environments should be distinct.

In this paper, we compare the ISM properties of GRB 060505's host galaxy
to long and short GRB hosts, as well as a large sample of blue compact galaxies.  We show that GRB 060505 has a unique set of ISM properties that leads to insight into the puzzling nature of its progenitor.
Throughout this paper we assume a cosmology of $H_0$ = 70 km s$^{-1}$ Mpc$^{-1}$, $\Omega_m$ = 0.3, and
$\Omega_{\Lambda}$ = 0.7.

\section{Emission Line Fluxes}
We use emission line fluxes from Th\"{o}ne et al. (2007) for five different regions of the host galaxy: (a)
the site of the gamma-ray burst, (b) the upper, (c) middle, and (d) lower regions of the
galaxy's bulge, and (e) a region of the galaxy's lower spiral arm. 

Our comparison sample is composed
of sixty-seven blue compact galaxies (BCGs) from the spectroscopic study of Kong \& Cheng (2002), and seven long- and two
short- duration GRB host galaxies from the GHostS public archive (Savaglio et al. 2006) . The reference sources for these fluxes are given in Table \ref{tab:params}.

All LGRB fluxes were corrected for local extinction effects based on the H$\alpha$/H$\beta$
emission line ratio where possible.   We use the Cardelli, Clayton \& Mathis (1989)
reddening curve, assuming an ${\rm R_{V}=Av/{\rm E}(B-V)} = 3.1$ and an 
intrinsic H$\alpha$/H$\beta$ ratio of 2.85 (the Balmer decrement for case B 
recombination at T$=10^4$K and $n_{e} \sim 10^2 - 10^4 {\rm cm}^{-3}$;
Osterbrock 1989).   The E($B-V$) values applied are given in Table \ref{tab:params}. 

 H$\alpha$ line fluxes were unavailable for the short GRB hosts.
In these cases we estimate the extinction using the E($B-V$)-$M_B$ relation in Jansen et al. (2001). We use $M_B$ for GRB 051221 and GRB 050416 from Soderberg et al. (2006, 2007), 
respectively.  The resulting extinction values are $\sim 0.3$, consistent with the mean extinction for the Nearby Field Galaxy Survey (Kewley, Jansen \& Geller 2005), and with the mean extinction of star-forming galaxies in the Sloan Digital Sky Survey (Kewley et al. 2007).

\section{Line Ratio Diagnostics}
In figures~\ref{fig:N2Ha} and \ref{fig:N2O2} we show the common line diagnostic diagrams proposed
by Veilleux \& Osterbrock
(1987), Baldwin et al. (1981), and Dopita et al. (2000). For comparison, we show representative Mappings
photoionization grids from
Kewley et al. (2001). These grids model a continuous star
formation history, using the Starburst99 v3.0 stellar population synthesis models and the 
Mappings III photoionization models with an electron
density of 350 cm$^{-3}$ and an age of 4 Myr. The grids shown here
have ionization parameters ranging from $q = 5 \times 10^6$ to $3 \times 10^8$ cm s$^{-1}$ and
metallicities ranging from $Z = 0.09Z_{\odot}$ to $Z = 3.5Z_{\odot}$, where solar metallicity
is log(O/H) + 12 = 8.7 (Allende Prieto, Lambert, \& Asplund 2001). We discuss each diagram separately below.

\subsection{[NII]/H$\alpha$ vs. [OIII]/H$\beta$}
\begin{figure}
\epsscale{1.0}
\plotone{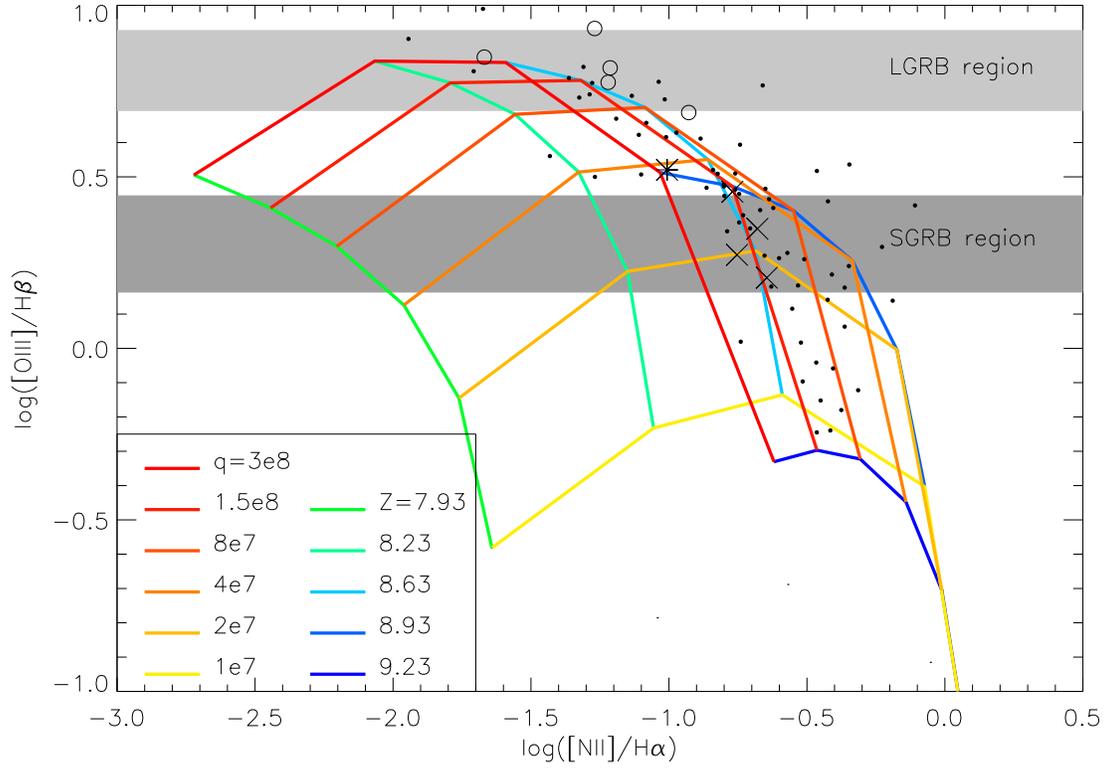}
\caption{\label{fig:N2Ha} Comparison of the [NII]/H$\alpha$ vs. [OIII]/H$\beta$
diagnostic for the Kong \& Cheng (2002) sample of BCGs (small points), the
long GRB host galaxies (open circles), the GRB region of GRB 060505's host (star), and
several bulge and spiral arm regions of GRB 060505's host (X's). The range of
ionization parameters delineated by the location of the long and short GRB hosts are shaded
and labeled as shown. The
galaxy positions are compared to the photoionization grids
of Kewley et al. (2001).}
\end{figure}

The [NII]/H$\alpha$ ratio is correlated strongly with metallicity and ionization parameter (Veilleux \&
Osterbrock 1987, Kewley \& Dopita 2002). The [OIII]/H$\beta$ ratio is sensitive to
 the ionization parameter of a galaxy and the hardness of the ionizing radiation field 
(Baldwin et al. 1981). From
Figure~\ref{fig:N2Ha}
we can see that the long GRB host galaxies are concentrated in the upper region of
the diagnostic diagram (0.688 $<$ log([OIII]/H$\beta$)$_{LGRBs}$ $<$ 0.932), with the short GRB hosts
in the lower region of the diagram (0.161 $<$ log([OIII]/H$\beta$)$_{SGRBs}$ $<$ 0.452). 
The BCG sample spans a much broader range of
[OIII]/H$\beta$ ratios (-0.245 $<$ log([OIII]/H$\beta$)$_{BCGs}$ $<$ 0.990).

Examining the placement of GRB 060505's burst site on this diagram, we see that
the [OIII]/H$\beta$ ratio of the GRB site (log([OIII]/H$\beta$) = 0.520)
places it well $below$ the region delineated by
the long GRB hosts, lying closer to the region occupied by the short GRB hosts. Its
[NII]/H$\alpha$ ratio cannot distinguish it from long or short GRB hosts. The other regions
of GRB 060505's host galaxy all have higher [NII]/H$\alpha$ ratios and lower
[OIII]/H$\beta$ ratios than the long GRB hosts, occupying the same region as the short GRB hosts
in Figure~\ref{fig:N2Ha}.

\subsection{[NII]/[OII] vs. [OIII]/[OII]}
\begin{figure}
\epsscale{1.0}
\plotone{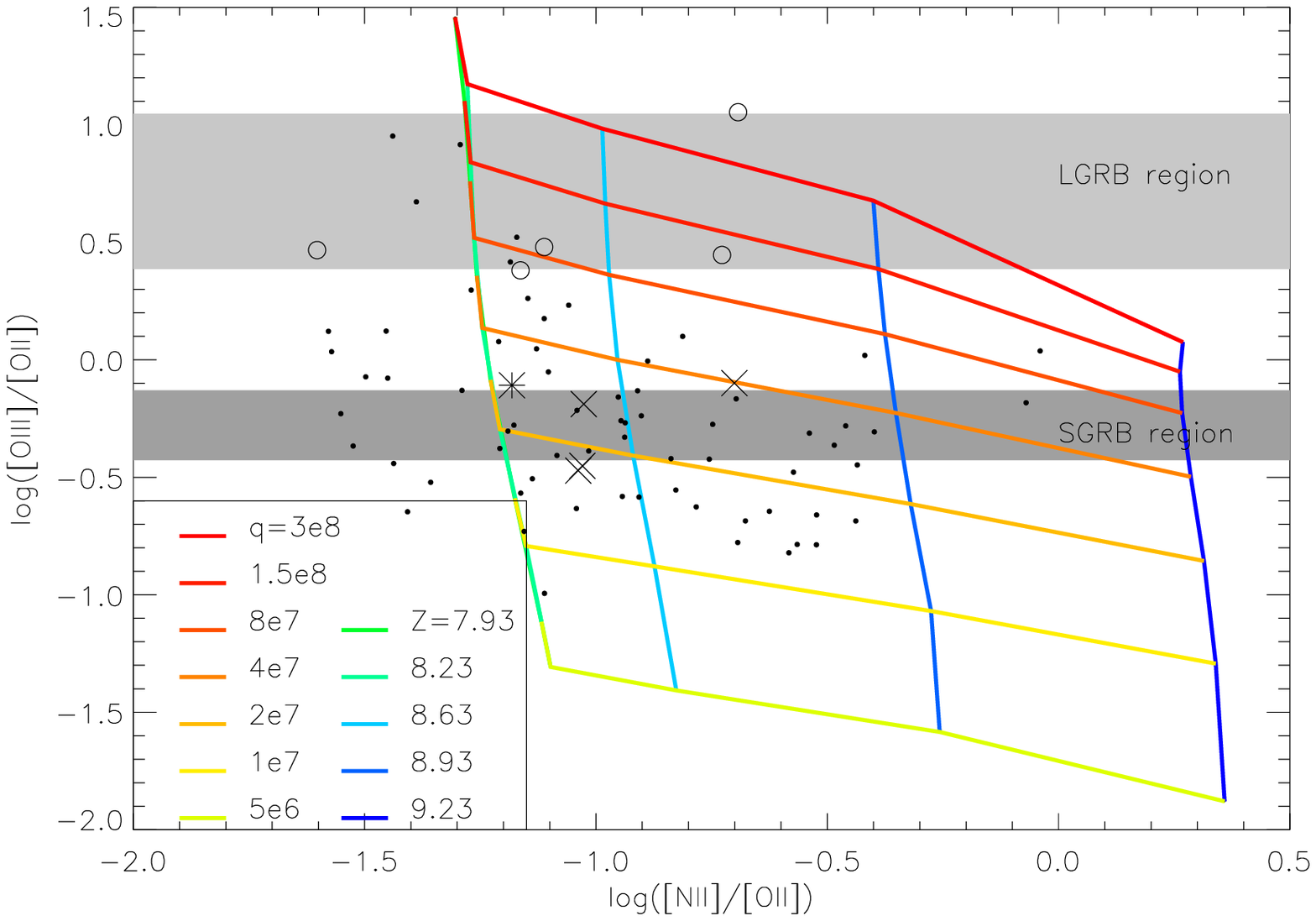}
\caption{\label{fig:N2O2} Comparison of the [NII]/[OII] vs. [OIII]/[OII]
diagnostic for the Kong \& Cheng (2002) sample of BCGs (points), the
long-duration GRB host galaxies (open circles), the GRB region of GRB 060505's host (star), and
several bulge and spiral arm regions of GRB 060505's host (X's). The range of
ionization parameters delineated by the location of the long and short GRB hosts are shaded
and labeled as shown. The
galaxy positions are compared to the photoionization grids
of Kewley et al. (2001).}
\end{figure}

As described by Baldwin et al. (1981), the [NII] and [OII] fluxes are directly
proportional to a galaxy's high-ionization volume, while the [OIII] flux
is directly proportional to the low-ionization volume. This makes the [NII]/[OII] vs.
[OIII]/[OII] diagnostic (Figure~\ref{fig:N2O2}) a powerful means of measuring a galaxy's
ionization parameter, with [NII]/[OII] primarily sensitive to metallicity and [OIII]/[OII] primarily
sensitive to ionization parameter.

In Figure~\ref{fig:N2O2} there is a large separation between the [OIII]/[OII] ratios (or ionization
parameters) of the host galaxies; long GRBs (0.380 $<$ log([OIII]/[OII])$_{LGRBs}$ $<$ 1.053) and short GRBs
(-0.409 $<$ log([OIII]/[OII])$_{SGRBs}$ $<$ -0.108). The BCG sample spans a much larger
range (-0.994 $<$ log([OIII]/[OII])$_{BCGs}$ $<$ 0.952).

The [OIII]/[OII] ratio of the GRB 060505 burst site (log([OIII]/[OII]) = -0.108) lies well below (0.5 dex)
the region
of the long GRBs, instead resting at the upper limit of the short GRB region. The [NII]/[OII] ratio of the burst
site agrees well with the long GRB hosts and BCGs. The other regions of GRB 060505's host galaxy show
similarly intermediate [NII]/[OII] ratios and [OIII]/[OII] that are consistent with the short GRB host region.

The metallicity-sensitive [NII]/[OII] ratio suggests that the GRB 060505 burst site has a
marginally lower metallicity than the other regions of its host galaxy, while the
[OIII]/[OII] ratio shows that the ionization parameter of the GRB 060505 burst site is consistent with the
short GRB region of the diagram. Figures~\ref{fig:N2Ha} and \ref{fig:N2O2}
indicate that the ISM properties of both the GRB 060505 burst site and the other
regions of the galaxy are consistent with the ISM properties of short-duration GRBs associated with compact-object mergers.

\section{ISM Properties}
We calculate metallicities using the R$_{23}$ diagnostic originally described
in Kewley \& Dopita (2002) and later refined and quantified in Kobulnicky \& Kewley
(2004). The R$_{23}$ diagnostic is double-valued, so we use the [NII]/[OII] criterion of Kewley 
\& Dopita (2002) to distinguish between the upper (log([NII]/[OII]) $>$ -1.2 and lower
(log([NII]/[OII]) $<$ -1.2) R$_{23}$ diagnostics. The [NII]/[OII] ratio is available
for five of our long GRB host galaxies as well as all five
regions of the GRB 060505 host galaxy and the entire BCG sample. In the absence of 
the [NII] $\lambda$6584 flux (GRB 010921 and GRB 990712), we distinguish between the branches using the
[NeIII]/H$\alpha$ ratio (Nagao et al. 2006). When
neither [NII] nor [NeIII] is available (GRB 051121 and GRB 050416),
we calculate the metallicities for both R$_{23}$ branches.
For comparison, we also calculate the metallicities using the Pettini \& Pagel (2004)
[NII]/H$\alpha$-metallicity relation for our five long GRB hosts with [NII]$\lambda$6583 line
fluxes.

We determine the ionization
parameter using the Kewley \& Dopita (2002) [OIII]/[OII]-$q$ relation.
For the five regions of GRB 060505's host galaxy, we also calculate the age of the
young ($<$ 10 Myr) stellar population using the calibration of H$\beta$ equivalent width with age by
Schaerer \& Vacca (1998). We calculate the H$\alpha$ luminosities and corresponding star formation rates (SFRs)
using the relation of Kennicutt (1998)
for the GRB 060505 host regions and our long GRB hosts (H$\alpha$ fluxes were unavailable for the short
GRB hosts). The metallicities, ionization parameters, H$\alpha$ luminosities, and SFRs are given in Table~\ref{tab:params}.

We find that the GRB 060505 site has a similar metallicity to
the other regions of its host galaxy; the R$_{23}$ diagnostic assigns the
GRB 060505 burst site an intermediate metallicity of log(O/H) + 12 = 8.57 $\pm$ 0.01 as compared to the rest of the host
galaxy, with an average log(O/H) + 12 = 8.62 $\pm$ 0.2.
The Pettini \& Pagel (2004), or PP04, relation gives the GRB 060505 site a lower metallicity of log(O/H) + 12 = 8.28 $\pm$ 0.01
than the average metallicity of the other regions (log(O/H) + 12 = 8.45 $\pm$ 0.06), but the difference is within the 0.15 dex
errors of the PP04 method. The relative metallicity of the GRB 060505 site to the other regions of the galaxy is dependent on the diagnostic that is used; when the errors of these diagnostics are considered (0.1-0.15~dex), we find no statistically
significant difference in metallicity between these five regions.

The ionization parameter of the GRB 060505
site, log($q$) = 7.49~cm/s, is unusually low as compared with average ionization parameter of the long GRB host galaxies
(log($q$) = 7.95 $\pm$ 0.28~cm/s. The ionization parameter of the GRB 060505 site is consistent with the short GRB hosts,
which have an average ionization parameter of log($q$) = 7.36 $\pm$ 0.15~cm/s.

The ages of the young stellar
populations are representative of core-collapse progenitor ages
(see Bloom et al. 2002, Berger et al. 2007). While the GRB site's is the youngest at 5.3 $\pm$ 0.3,
this age is comparable to the 6.1 $\pm$ 0.6 age of the young stellar population of the
upper bulge, to within the errors.

Finally, the GRB site does not have a 
high SFR with respect to the other regions of the host galaxy. The SFR of the GRB site
is also found to be notably lower than the SFRs of the long GRB host galaxies.

\section{Conclusions}
We compare the optical diagnostic emission-line ratios of GRB 060505 with the hosts of unambiguous long- and short- duration GRBs.\
We show that the emission-line ratios, metallicities, ionization parameters and star formation rates of the GRB 060505 environment are more consistent with the two short-duration GRB hosts that have measured optical emission-line ratios rhan with long-duration GRBs.

We compare the metallicity and SFR of the GRB 060505 star-forming region with four other star-forming regions in the GRB 060505 host galaxy.  We find no significant difference in either metallicity or SFR between the GRB 060505 region or the other star-forming regions, including the host galaxy bulge.  

 We do not find compelling evidence to suggest that GRB 060505 originated in a long-duration core-collapse progenitor.    Our emission line diagnostic analysis suggests that the environment of GRB 060505 is more consistent with the host environments of compact-object-merger GRB progenitors.  A larger comparison sample of short and long GRBs with emission line spectra may shed further light on the nature of  GRB 060505 and other  intermediate-duration gamma-ray bursts.

\clearpage
\begin{deluxetable}{l l l l l l l l l l l}
\tabletypesize{\scriptsize}
\tablewidth{0pc}
\tablenum{1}
\tablecolumns{11}
\tablecaption{\label{tab:params} ISM Properties of GRB Host Galaxies}
\tablehead{
\colhead{Galaxy}
&\colhead{$z$\tablenotemark{a}}
&\multicolumn{2}{c}{log(O/H) + 12\tablenotemark{b}}
&\colhead{log($q$)}
&\colhead{E($B-V$)}
&\colhead{W$_{H\beta}$}
&\colhead{Age (Myr)\tablenotemark{c}}
&\colhead{L(H$\alpha$) (ergs s$^{-1}$)}
&\colhead{SFR (M$_{\odot}$/yr)}
&\colhead{Refs.\tablenotemark{d}} \\ \cline{3-4}
\multicolumn{2}{c}{}
&\colhead{R$_{23}$}
&\colhead{PP04}
&\multicolumn{7}{c}{}
}
\startdata
GRB 060505 HOST & & & & & & & & & & \\
\hline
GRB region       &0.0889 &8.57$\pm$0.01 &8.28$\pm$0.01 &7.49 &0.13 &39.20 &5.1$\pm$0.2 &$2.61 \times 10^{39}$ &0.021 &1\\
Upper bulge      &0.0889 &8.84$\pm$0.01 &8.47$\pm$0.01 &7.66 &$<$0.03 &27.22 &5.4$\pm$0.6 &$2.38 \times 10^{39}$ &0.019 &1\\
Middle bulge     &0.0889 &8.43$\pm$0.00 &8.50$\pm$0.00 &7.09 &0.59 &8.06  &7.8$\pm$0.4 &$1.69 \times 10^{40}$ &0.134 &1\\
Lower bulge      &0.0889 &8.59$\pm$0.01 &8.42$\pm$0.00 &7.25 &0.30 &12.60 &6.7$\pm$0.3 &$5.09 \times 10^{39}$ &0.040 &1\\
Lower spiral     &0.0889 &8.60$\pm$0.06 &8.41$\pm$0.04 &7.44 &$<$0.03 &8.92  &7.5$\pm$0.4 &$3.12 \times 10^{38}$ &0.002 &1\\
\hline
LONG GRBs & & & & & & & & & & \\
\hline
GRB 060218 &0.0335 &8.35$\pm$0.01 &8.19$\pm$0.02 &7.78 &0.17 &\nodata &\nodata &$1.22 \times 10^{40}$ &0.096 &2,3\\
GRB 031203 &0.1055 &8.26$\pm$0.01 &8.17$\pm$0.01 &8.32 &0.03 &90.72 &4.7 $\pm$ 0.1 &$1.51 \times 10^{42}$ &11.93 &4\\
GRB 030329 &0.1680 &8.72$\pm$0.03 &8.32$\pm$0.05 &8.08 &$<$0.03 &\nodata &\nodata &$2.54 \times 10^{40}$ &0.201 &5,6\\
GRB 020903 &0.2510 &8.15\tablenotemark{e} &8.00 &7.88 &0.25 &36.08 &5.5 $\pm$ 0.2 &$2.57 \times 10^{41}$ &2.030 &7\\
GRB 010921 &0.4510 &8.55$\pm$0.03 &\nodata &7.67 &0.52 &\nodata &\nodata &$8.21 \times 10^{41}$ &6.486 &8\\
GRB 990712 &0.4340 &8.43$\pm$0.01 &\nodata &7.74 &0.17 &\nodata &\nodata &$4.63 \times 10^{41}$ &3.658 &9,10\\
GRB 980425 &0.0085 &8.36\tablenotemark{e} &8.19 &7.86 &0.60 &\nodata &\nodata &$1.20 \times 10^{40}$ &0.095 &11\\
\hline
SHORT GRBS & & & & & & & & & & \\
\hline
GRB 051221 &0.5464 &8.25 or 8.80 &\nodata &7.17 or 7.36 &0.34\tablenotemark{f} &\nodata &\nodata &\nodata &\nodata &12\\
GRB 050416 &0.6528 &8.30 or 8.68 &\nodata &7.38 or 7.53 &0.34\tablenotemark{f} &\nodata &\nodata &\nodata &\nodata &13\\

\enddata
\tablenotetext{a}{Redshifts come from the GHostS database.}
\tablenotetext{b}{Errors are propagated from statistical flux errors. These do not include the systematic
error introduced by the metallicity calibrations, which are 0.1 dex for the R$_{23}$ metallicities (Kewley
\& Dopita 2002) and 0.15 dex for the PP04 metallicities (Pettini \& Pagel 2004).}
\tablenotetext{c}{Ages comes from equations derived for the  Schaerer \& Vacca (1998) 
models relating H$\beta$ equivalent widths ($W_{H\beta}$) and the age of the young stellar population.}
\tablenotetext{d}{Reference: (1) Th\"{o}ne et al. (2007), (2) Pian et al. (2006), (3) Wiersema et al. (2007),
(4) Prochaska et al. (2004), (5) Gorosabel et al. (2005), (6) Hjorth et al. (2003), (7) Soderberg et al. (2004),
(8) Price et al. (2002), (9) K\"{u}pc\"{u} Yoldas et al. (2006), (10) Christensen et al. (2004), (11) Hammer et al. (2006),
(12) Soderberg et al. (2006), (13) Soderberg et al. (2007)}
\tablenotetext{e}{Statistical flux errors not available in literature.}
\tablenotetext{f}{Derived from the $M_B$-E($B-V$) relation of Jansen et al. (2001).}
\end{deluxetable}

\clearpage

\end{document}